\documentstyle[12pt]{article}
\title{Calculation of a weak nonleptonic matrix\\
element using ``Weinberg'' sum rules}
\author{John F. Donoghue\\[5mm]
Department of Physics and Astronomy\\
University of
Amherst, MA ~01003}
\date{}
\begin{document}
\begin{titlepage}
\maketitle
\begin{abstract}

There is a ``toy'' weak matrix element which can be expressed as
an integral over the  vector and axial vector spectral functions, $\rho_V (s) -
\rho_A (s)$.  I review our recent  evaluation of these spectral functions, the
study of four ``Weinberg'' sum rules and the calculation  of this matrix
element. [Talk presented at the XXVIII International Conference on High Energy
Physics, ICHEP94, Glasgow, Aug. 1994, to be published in the proceedings.]

\end{abstract}
{\vfill UMHEP-412 \\ hep-ph/9409398}
\end{titlepage}

Weak nonleptonic matrix elements are notoriously difficult to calculate by any
reliable method.  E. Golowich and I have proposed a novel weak matrix element,
not found in the standard model, which can be well calculated by a mixture of
theoretical and phenomenological inputs[1].  In the process we had to update
the
status of the Weinberg sum rules[2].  This
talk briefly reviews these developments.

Consider a weak matrix element formed using only vector currents

\begin{eqnarray}
\tilde{\cal H}_w = {g^2 \over 8} \int d^4x D_F (x, M_w)
T\left( \bar{d}(x)
\gamma_{\mu} u (x)
\bar{u} (0) \gamma^{\mu} s (0) \right).
\end{eqnarray}

\noindent Aside from KM factors, this differs from the weak Hamiltonian of the
Standard Model
only in that the latter involves left handed (V-A) currents.
However under chiral symmetry the Hamiltonian in Eq. 1
has a different transformation property since $8_V = 8_L \oplus 8_R$,
where V, L, R refer to
transformations under vectorial, left handed and right handed SU(3)
respectively.
The $8 \oplus 8$
direct product contains an $(8_L, 8_R)$ term

\begin{eqnarray}
8 \otimes 8 &=& (8_L, 8_R) + (8_L, 1_R) \nonumber \\
&+& (1_L, 8_R) + (27_L, 1_R) + (1_L, 27_R).
\end{eqnarray}

The  $(8_L, 8_R)$ component is special because it is the
only term which does not
vanish in the
chiral limit $(m_q \rightarrow 0, p \rightarrow 0)$.
The value of a $K \rightarrow
\pi$ matrix
element in the chiral limit (which
we adopt hence forth) is then calculable using the
soft pion
theorem to remove the pseudoscalars

\begin{eqnarray}
\lefteqn{ \langle \pi (p) \mid \tilde{\cal{H}}_w \mid K(p) \rangle }
\nonumber \\
& & ={-g^2 \over F^2_{\pi}} \int d^4x D_F (x, M_w) \nonumber \\
& & \hspace{ .5in}\langle 0 \mid T
 \left( V_{\mu} (x) V^{\mu} (0) -
A_{\mu} (x) A^{\mu}(0) \right) \mid 0 \rangle \nonumber \\
& & = {-g^2 \over F^2_{\pi}} \int d^4x D_F (x, M_w) \left[ \pi_V (x)
 - \pi_A (x) \right]
\end{eqnarray}

\noindent where $\pi_{V,A}$ are the
vector and axial polarization tensors.  After
writing these in
terms of the spectral densities, one obtains

\begin{eqnarray}
\langle \pi (p) \mid \tilde{\cal{H}}_w \mid K(p) \rangle = {G_F
\over \sqrt{2}} A
\end{eqnarray}

\noindent where

\begin{eqnarray}
A = M^2_w \int ds {s^2 \over s - M^2_w} ln \left( s / M^2_w \right)
\left[ \rho_V (s)
-  \rho_A (s) \right]
\end{eqnarray}

\noindent This is similar to four other sum rules.

\begin{eqnarray}
\int {ds \over s} \left[ \rho_V (s) - \rho_A (s) \right]
= -4 \bar{L}_{10} \nonumber
\\
\int ds \left[ \rho_V (s) - \rho_A (s) \right] = F^2_\pi \nonumber \\
\int ds~ s \left[ \rho_V (s) - \rho_A (s) \right] = 0 \nonumber \\
\int ds~ s~ ln(s) \left[ \rho_V (s) - \rho_A (s) \right] \nonumber \\
= -{16\pi^2 F^2_\pi \over 3 e^2}\left( m^2_{\pi} -
m^2_{\pi}  \right)
\end{eqnarray}

\noindent valid in the chiral limit. Here the second and third of these are the
original two Weinberg sum rules[3]. The first sum rule above involves the
chiral
coefficient $\bar{L}_{10}=-(9.1 \pm 0.3) \times 10^{-3}$ which is measured
in radiative pion
decay. This sum rule originates in work of Das et al and was given in its
present, more general, form by Gasser and Leutwyler[4]. The final sum rule
comes from the calculation of the electromagnetic contribution to the
mass difference of neutral and charged pions[5] at lowest order in
the chiral expansion. Although these were first derived before QCD, they
rely on assumptions about the short distance properties that can only be
proven through the use of QCD in the chiral limit.
Note that the last two sum rules are no longer
valid if the quark masses are turned on. This set of sum rules represents a
beautiful interplay of the chiral and short distance properties of
QCD.

The spectral functions can be constructed
fairly reliably.  This is not the place to
discuss all aspects
[see Ref. 2], but the low energy portion is
known from chiral symmetry and the high
energy effects are
small and amenable to treatment by perturbative
QCD.  The intermediate energy contributions are
not
theoretically calculable at present, but fortunately these
may be extracted from $e^+
e$ and $\tau$
decay data.  The vector spectral function starts out with two-pion and
four-pion contributions, and relatively quickly approaches a constant value.
The axial spectral function has three and five pion contributions and
approaches
the same constant. The difference between them goes to zero as $s^{-3}$, which
vanishes so rapidly that there is not much contribution to the sum rules from
high energy.
There are minor uncertainties in the data,
and we adjust the spectral
functions within the range of experimental uncertainties in order to fit the
data while accommodating the four Weinberg sum
rules.  The resulting forms for
$\rho_V$ and
$\rho_A$ separately and for the difference are given in Ref 2.  While this
procedure does not prove the Weinberg sum rules are required by
the data, they certainly
are easily compatible with the set of experimental information. The fact
that it is easy to satisfy the Weinberg sum
rules within the
constraints of theory and data is very nontrivial and is a credit to the
complex theoretical ideas that went into their formulation.

When applied to the weak matrix element we obtain

\begin{eqnarray}
A =-0.062 \pm 0.017 GeV^6\nonumber \\
\langle \pi (p) \mid \tilde{\cal{H}}_w \mid K(p) \rangle = 5.3 \times 10^{-7}
GeV^2
\end{eqnarray}

\noindent In contrast, the ``vacuum saturation''
approximation would yield
\begin{eqnarray}
A_{vac-sat} = -0.033 GeV^6
\end{eqnarray}
and the
real weak matrix
element extracted from $K \rightarrow 2 \pi$ using chiral symmetry

\begin{equation}
{\langle \pi (p) \mid {\cal H}_w \mid K (p) \rangle
\over \mid V_{ud} V^{\ast}_{us}
\mid} = 1.7 \times 10^{-7} GeV^2
\end{equation}

\noindent We see a modest enhancement of the matrix element.

This calculation has not uncovered the mechanism for
the $\Delta I = {1 \over 2}$
rule, as the
$(8_L, 8_R)$  operator automatically does not have the
freedom to have a $\Delta I
= {1 \over 2}$
enhancement, requiring $A_{3/2} = {2 \over 3} A_{1/2}$ always.  However
inspection of the
details of the calculation does reveal a
hint as to why it is so difficult to calculate
nonleptonic
amplitudes.  There is very little contribution
from either the high or low energy
ends, where theory
is useful.  Most of the strength comes
from intermediate energies, which are generally not under
theoretical control.
While we cannot apply this matrix element to
Standard Model phenomenology, it
should prove
possible to use it as a test of lattice calculational methods. In
addition, it is possible that this calculational technique may be extended to
study more realistic matrix elements.

%\section{References}

\vspace{5mm}
\noindent
\end{document}